\documentclass[a4paper]{jpconf}
\usepackage{graphicx}
\usepackage[square,sort&compress]{natbib}
\usepackage{aas_macros}
\begin{document}

\title{Local helioseismology of sunspot regions: comparison of ring-diagram and time-distance results}

\author{A.G.~Kosovichev$^1$,   S.~Basu$^2$, R.~Bogart$^1$, T.L.~Duvall,~Jr$^3$, I.~Gonzalez-Hernandez$^4$, D.~Haber$^5$, T.~Hartlep$^1$, R.~Howe$^4$,  R.~Komm$^4$, S.~Kholikov$^4$, K.V.~Parchevsky$^1$, S.~Tripathy$^4$, J.~Zhao$^1$
}

\address{$^1$~Stanford University, Stanford, CA, USA; $^2$Yale University, New Haven, CT  06520, USA; $^3$Solar Physics Laboratory, Goddard Space Fight Center, NASA, Greenbelt, MD 20771, USA; $^4$National Solar Observatory, Tucson, AZ 85719, USA; $^5$JILA, University of Colorado, Boulder, CO 80309, USA}

\ead{AKosovichev@solar.stanford.edu}

\begin{abstract}
Local helioseismology provides unique information about the
subsurface structure and dynamics of sunspots and active regions.
However, because of complexity of sunspot regions
local helioseismology diagnostics require careful
analysis of systematic uncertainties and physical
interpretation of the inversion results. We present new results
of comparison of the ring-diagram analysis and time-distance
helioseismology for active region NOAA 9787, for which a previous
comparison showed significant differences in the subsurface
sound-speed structure, and discuss systematic uncertainties
of the measurements and inversions. Our results show that both the ring-diagram and time-distance techniques give qualitatively similar results, revealing a characteristic two-layer seismic sound-speed structure consistent with the results for other active regions.
However, a quantitative comparison of the inversion results is not straightforward. It must take into account differences in the sensitivity, spatial resolution and the averaging kernels. In particular, because of the acoustic power suppression, the contribution of the sunspot seismic structure to the ring-diagram signal can be substantially reduced. We show that taking into account this effect reduces the difference in the depth of transition between the negative and positive sound-speed variations  inferred by these methods. Further detailed analysis of the sensitivity, resolution and averaging properties of the local helioseismology methods is necessary for consolidation of the inversion results. It seems to be important that both methods indicate that the seismic structure of sunspots is rather deep and extends to at least 20 Mm below the surface, putting constraints on theoretical models of sunspots.
\end{abstract}

\section{Introduction: sunspot models and helioseismology}
Sunspots and active regions are key elements of solar magnetism. Solar magnetic field is generated by a dynamo process in the interior. It appears on the surface in the form of highly concentrated magnetic structures, sunspots, often surrounded by plages representing large areas of diffuse magnetic field. The mechanism of formation and stability of sunspots is currently not understood. Observations of the solar surface provide evidence that sunspots are formed by merging together small-scale magnetic elements, which appear on the surface as emerging magnetic flux. After sunspots are formed, they may be stable for several weeks. Long-living sunspots are often accompanied by new emerging magnetic flux events. The decay of sunspots is associated with a rapid diffusive process in the form of enhanced moat flow and magnetic field submergence. It is quite clear that all these processes are controlled by subsurface dynamics, but how turbulent convection can lead to formation of such stable self-organized magnetic structures is a great puzzle.

Recent numerical simulations of magnetic structures on the Sun lead us to two basic theoretical ideas about the sunspot formation and stability. The first idea comes from the simulations of Rempel et al \cite{Rempel2009a}, in which the sunspot structure is supported by fixing a localized magnetic field concentration at the bottom boundary. These simulations show formation of mean converging downflows around a pore-like structure, which does not have a penumbra. When the penumbra-like magnetic structure is modeled (by setting up appropriate upper boundary conditions) then the simulations show that the subsurface dynamics is dominated by outflows, similar to the Evershed effect. These flows are driven by the overturning convection and Lorentz force near the surface, but the diverging flows extend through the whole depth of the computational domain \cite{Rempel2010}. A possible explanation for the deep diverging upflows is that the penumbra partially block the turbulent convective heat flux, and it needs to be transported by upflows (Rempel, private communication). In both cases, once the bottom boundary condition is released the structure disappears on the time scale of the convective turn-over time. Thus, for long-term stability in this model, it is necessary that the sunspots are formed by strong-field flux tubes anchored in the deep interior, where the convective turn-over time is large, and therefore, the lifetime of sunspots corresponds to this turn-over time.

Another important idea comes from the simulations of spontaneous formation of magnetic structures from an initially uniform magnetic field, recently obtained by Kitiashvili et al \cite{Kitiashvili2010}. In this model, a stable pore-like structure (without penumbra) is formed by merging small-scale magnetic elements (flux tubes) initially concentrated by vortexes (whirlpool-like structures) in the intergranular lanes. The magnetic field strength in this self-organized magnetic structure is about 1.5 kG at the surface and 6 kG in the interior. The boundary conditions in these simulations only keep the mean magnetic field strength constant in the simulation domain, and no artificial boundary conditions are used to maintain the structure. It is intrinsically self-maintained.  The principal mechanism of the structure formation and stability is associated with strong converging downdrafts. The sunspot penumbra has not been simulated for the self-organized structure model, but a separate computation of magnetoconvection in highly inclined strong magnetic fields shows that the Evershed flows are likely to be confined in the top 1Mm deep layer \cite{Kitiashvili2010a,Kitiashvili2009}. The magnetic pore-like structure has an internal cluster-type structurization. The simulations indicate that for formation of a large sunspot-like structure it is necessary to increase the depth of the simulation domain.

These two radiative MHD simulations provide nice examples of the classical Parker's sunspot dilemma: monolithic vs cluster model \cite{Parker1979}. The Rempel's model belongs to the monolithic type while Kitiashvili's simulations give a demonstration of a cluster-type model. Observations of sunspots on the solar surface cannot resolve this dilemma. It seems that the process of sunspot formation and the flow dynamics dominated by converging flows is similar to the cluster model until penumbra is developed \cite{VargasDominguez2010,VargasDominguez2008}. Both MHD models strongly suggest that the sunspot structure and flows extend into the deep interior because of the long-term stability requirement. But so far, the models have been  calculated only for relatively shallow computational domains (6-8 Mm deep) due to the current computing power limitations.

In addition to the MHD sunspot models, several magnetostatic models were proposed (e.g. \cite{Cameron2010,Khomenko2008}). These models do not include flows and are calculated using the hydrostatic pressure balance for prescribed parametric distributions of magnetic field and density or temperature. The main goal of these models is to provide background models for linear wave simulations, used for testing local helioseismology inferences. These models can be quite shallow in terms of the relative seismic perturbations.

The surface structure and dynamics of sunspots have been studied in detail by high-resolution imaging and spectro-polarimetric measurements. Studying the interior properties can be done only by local helioseismology methods, and this is a very challenging task. Helioseismic inversions have to rely on our understanding of the oscillation physics in a very complicated environment of the highly turbulent magnetized plasma. In addition, they represent a classical ill-posed problem, which does not have a unique solution. Also, for computational efficiency the relationship between the observational data and interior properties is linearized. In the case of strong perturbations, when more accuracy is required, an iteration scheme can be applied. The advantage of this approach that it allows to investigate subsurface structures and dynamics without a priori model constraints (usually, only with smoothness constraints), and also to explore the sensitivity and resolution of helioseismology techniques. The accuracy of these techniques must be investigated by numerical simulations. Currently, there are several local helioseismology methods, including the ring-diagram analysis, acoustic holography and time-distance helioseismology. The current status of the local helioseismic diagnostics of sunspots was recently reviewed by Kosovichev \cite{Kosovichev2010}.

Our goal is to investigate and compare the inversion results for subsurface sound-speed variations of AR 9787, obtained by two different local helioseismology techniques, the ring-diagram analysis and time-distance helioseismology. This work was initiated by the LoHCo (Local Helioseismology Comparison) team to check the results of the HELAS group \cite{Gizon2009}. In this paper, we point out that the quantitative comparison of the ring-diagram and time-distance inversion results is not straightforward. It must take into account differences in the sensitivity and spatial resolution of these methods. In particular, the nonuniform distribution of the oscillation power in active region and differences in the averaging kernels of the inversions may have significant effects. Despite the differences, we find that both techniques indicate that sunspots and active regions are associated with sound-speed perturbations in the deep layers of the upper convection zone. These results are not consistent with "shallow" models of sunspots.

\section{Local helioseismology of AR 9787}
\subsection{Structure and evolution of active region}
\begin{figure}[h]
\begin{center}
\includegraphics[width=0.7\linewidth]{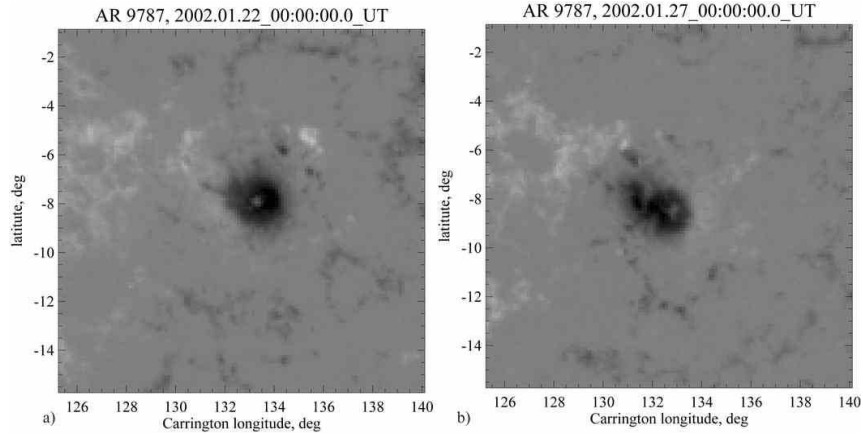}
\end{center}
\caption{\label{fig1}SOHO/MDI magnetogram of a 15-degree area around the leading sunspot of AR 9787, at two location approximately (a) 30 degrees West, and (b) 30 degrees East  of the central meridian. The magnetograms are remapped into the heliographic coordinates using the Postel's projection with the resolution of 0.12 deg/pixel. The gray scale corresponds to the magnetic field strength from -1500 G (black) to 1500 G (white).}
\end{figure}

\begin{figure}[h]
\begin{center}
\includegraphics[width=0.7\linewidth]{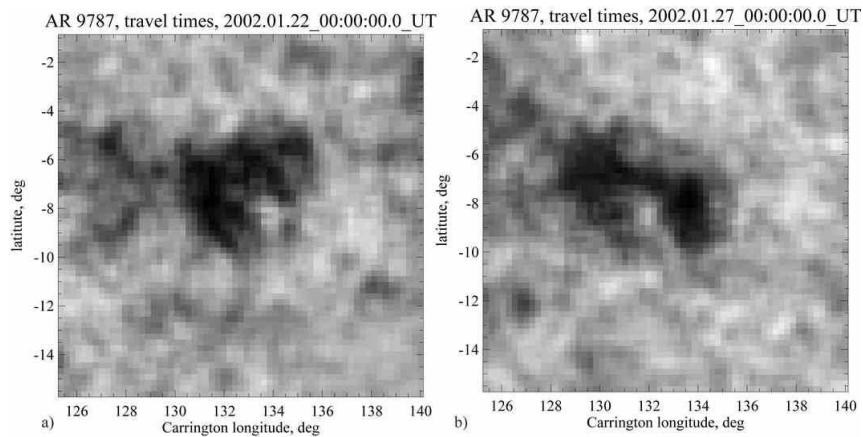}
\end{center}
\caption{\label{fig2}SOHO/MDI mean travel time of acoustic waves for the distance range of 1.38--1.86 degrees of the 15-degree areas of AR 9787 as in Fig.~\ref{fig1}. The gray scale corresponds to the travel times from 30.15 min (black) to 30.6 min (white). The travel times are calculated with the resolution of 0.24 deg/pixel and shown after 2x2-pixel smoothing.}
\end{figure}

Active region NOAA 9787 was observed by SOHO/MDI on January, 20-28, 2002. It had a complex magnetic structure, Beta-Gamma, according to the NOAA classification. Its location was at 6 degrees south latitude and 130 degrees Carrington longitude.  The leading sunspot had initially a round magnetic structure, but then it became significantly distorted (Fig.~\ref{fig1}), probably because of new emerging magnetic flux. The total magnetic flux during the interval of rotation between two symmetrical, relative to the central meridian, locations ($\pm 30^\circ$) increased by more than 20\%, from $7.9\times 10^{21}$ Mx to $9.7\times 10^{21}$ Mx. The leading sunspot was surrounded by a strong plage region with magnetic field reaching $\sim 750$ G. The plage region produced significant seismic travel-time anomalies comparable with the anomalies of the sunspot (Fig.~\ref{fig2}).

The travel-time anomaly in the sunspot region exhibits the sign reversal: the travel time perturbations relative to a quiet Sun region are positive for the short travel distances, in the range of 0.54-1.02 and 1.02-1.38 degrees, and negative for longer distances. In the plage region, the travel-time anomalies are negative for all distances. This fact has important implications for comparison of the ring-diagram and time-distance results.

\subsection{Ring-diagram analysis}

The ring-diagram analysis was carried out by using two fitting methods, described by Haber et al \cite{Haber2000} and Basu et al \cite{Basu2004}. These two techniques give similar results. We plan to discuss their comparison in a separate paper. Here we present the results obtained by the technique \cite{Basu2004}. The same technique was used for the analysis presented in \cite{Gizon2009}. The results of this paper were substantially different from the previous ring-diagram studies of other active regions \cite{Basu2004,Bogart2008}.
During our investigation, we found that the results \cite{Gizon2009} were, in fact, obtained not for the intended target, AR 9787, which was on the solar disk in January 2002, but for AR 9829 observed in February 2002 during the following Carrington rotation. The active region 9829 had the Carrington coordinates close to the coordinates of AR 9787. Presumably, AR 9829 was a remnant of the decaying AR 9787. When this region appeared on the East limb on February 16, 2002 (and was given the new NOAA number), the sunspot area was reduced by a factor of 10, from 400 to 40 millionths of the solar hemisphere. Two days later,  by February 18,  sunspots in this region completely disappeared. Only a plage area without sunspots remained during that Carrington rotation. Thus, the ring-diagram results published in \cite{Gizon2009} were mostly for the plage region without sunspots. This partly explains the difference from the previous ring-diagram results.
\begin{figure}[h]
\begin{center}
\includegraphics[width=0.7\linewidth]{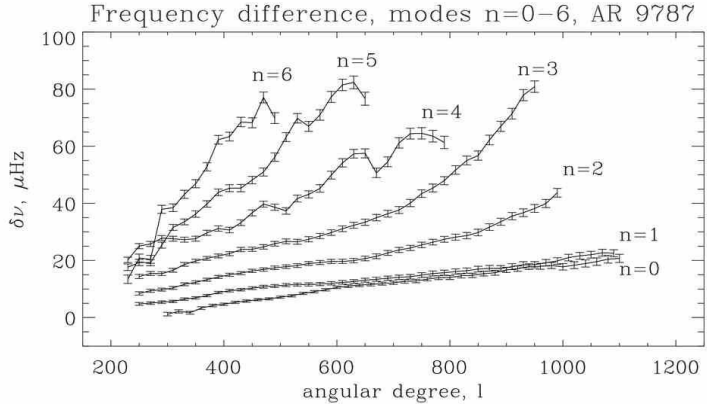}
\end{center}
\caption{\label{fig3}The frequency difference obtained by the ring-diagram technique for AR 9787 and a quiet-Sun region as a function of the angular degree, $l$, for the various radial order, $n$, of local oscillation modes.}
\end{figure}
\begin{figure}[h]
\begin{center}
\includegraphics[width=0.7\linewidth]{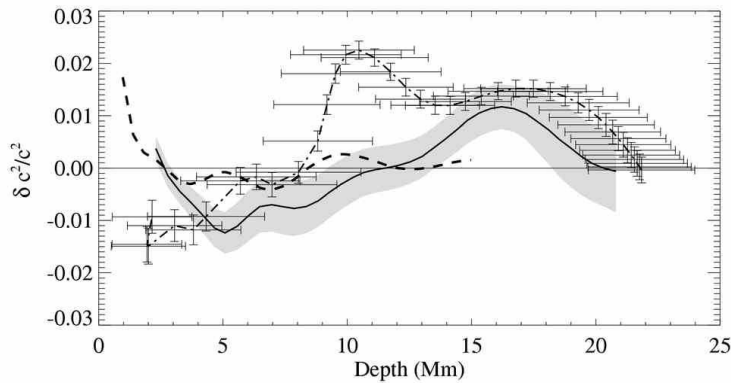}
\end{center}
\caption{\label{fig4}Comparison of the ring-diagram inversion results obtained for AR 9787 by two inversion technique: SOLA (solid curve with gray area indicating error estimates), and OLA (dash-dots with error bars), and the results of Gizon et al. \cite{Gizon2009} (dashed curve). The horizontal errorbars show an effective spatial resolution - spread of the averaging kernels, as defined by the inversion theory \cite{Kosovichev1999}.}
\end{figure}

We have carried out inversions of the frequency difference (Fig.~\ref{fig3}) measured by the ring-fitting technique \cite{Basu2004} for a 15-degree square region containing AR 9787, and a similar quiet-Sun region without sunspots. We used two independent inversion codes. Both codes are based on the variational principle for the frequency difference \cite{Gough1988}, but employ different inversion methods: Optimally Localized Averaging (OLA)  \cite{Kosovichev1999} and Subtractive Optimally Localized Averaging (SOLA)  \cite{Basu2004}.
In Figure \ref{fig4}, we show the inversion results of the frequency difference for the relative squared sound-speed perturbations obtained by these methods: OLA (dash-dots with errorbars), and SOLA (solid curve). For comparison we plot also the results published in \cite{Gizon2009} (dashed curve).
The main difference is that the results \cite{Gizon2009} showed a strong near-surface positive variation and very small variations in the deeper interior, while the new results show show mostly negative near-surface variations and significant  positive variations in the deep layers. The results obtained by both, the OLA and SOLA  inversion techniques, show a characteristic two-layer sound-speed structure. However, there are significant quantitative differences. The SOLA results show smaller variations, and the region of transition from the negative to positive variations appears shifted by 3-4 Mm down in depth. The reasons for these differences have to be investigated. They may be attributed to differences in the regularization (smoothness) parameters, which control the spatial resolution of the inversion results, in the averaging kernels and other effects.

\begin{figure}[h]
\begin{center}
\includegraphics[width=0.6\linewidth]{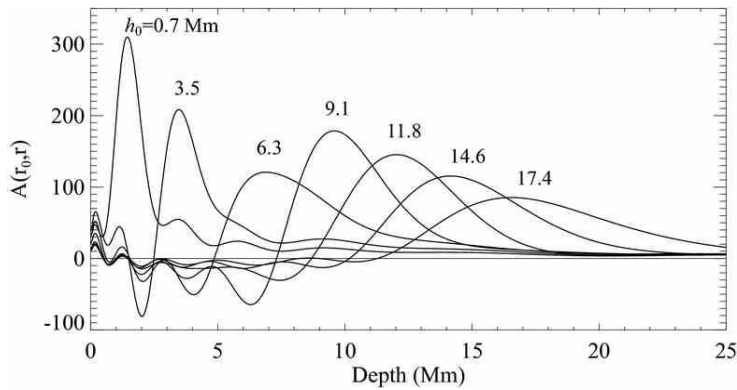}
\end{center}
\caption{\label{fig5}A sample of localized averaging kernels for the ring-diagram inversion. The values of $h_0$ are the corresponding target depths in Mm. }
\end{figure}

In general, it is important to realize that the inversion results do not give estimates precisely at a given depth, but represent a convolution with the averaging kernels. A sample of the averaging kernels obtained by the OLA method is illustrated in Fig.~\ref{fig5}. These kernels have small negative sidelobes and also are asymmetric. They show a good localization in depth between 1.5~Mm and 15~Mm. However, we were not able to obtain the good localization in the shallow and deeper layers. The averaging kernels of the SOLA method are shown in \cite{Basu2004}. They have a good localization in the same range of depth, and more regular shape, but appear somewhat broader in the near surface layers.

\begin{figure}[h]
\begin{center}
\includegraphics[width=0.7\linewidth]{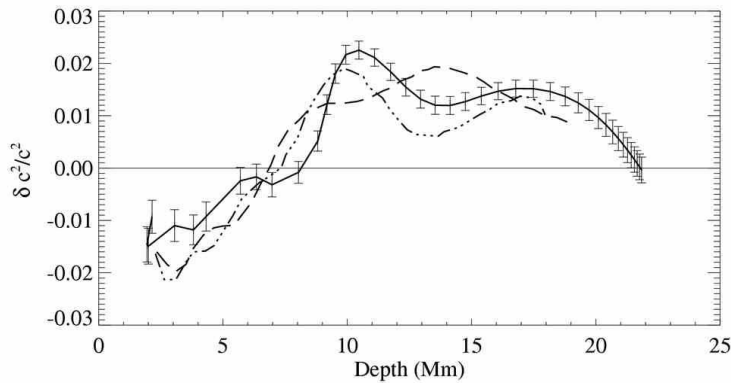}
\end{center}
\caption{\label{fig6}Comparison of the ring-diagram inversion results for AR 9787 obtained in this paper by the OLA inversion method (solid curve with error bars) with the inversion results of Bogart et al \cite{Bogart2008} two other active regions, AR 9906 (dashed curve) and AR 10793 (dot-dashed curve).}
\end{figure}

Our results for AR 9787 are in good agreement with the previous ring-diagram analysis of other active regions. A comparison with two active regions studied in \cite{Bogart2008} is shown in Fig.~\ref{fig6}. According to the statistical study of Baldner et al \cite{Baldner2009}, in most active regions the sound-speed variation obtained by the ring-diagram analysis has a negative variation near the surface and a positive variation in the deeper layer. Thus, the helioseismic structure of AR 9787 is not much different from other active regions.

\subsection{Comparison with results of time-distance helioseismology}

It has been shown by Basu et al \cite{Basu2004} and Bogart et al
\cite{Bogart2008} that this typical two-layer structure is qualitatively consistent with the time-distance inversion results
\cite{Kosovichev2000,Jensen2001,Jensen2003,Couvidat2006,Couvidat2006a,Zhao2010}. However,  the quantitative comparison is not straightforward because the ring-diagram and time-distance methods have quite different spatial and temporal resolutions. The time-distance method attempts to resolve structures close to the half-wavelength resolution limit (2-3 Mm), on the time scale of 8 hours, while the ring-diagram technique has a typical spatial resolution of about 180 Mm (15 heliographic degrees) and the time scale of 24 hours or longer. Thus, the regions analyzed by the ring-diagram method are much larger than typical sunspots (for illustration see Fig.~\ref{fig1}). Most of the oscillation signal in these areas comes not from sunspots, in which the oscillation power is suppressed, but from surrounding plage regions. The plage regions may have significant helioseismic effects (Fig.~\ref{fig2}), and thus they can make significant contribution to the mean seismic sound-speed profiles. These facts must be taken into account for quantitative comparisons between the ring-diagram and time-distance inversion results. In addition, as we have pointed out, the localization and spread of the averaging kernels also contribute to the inferred sound-speed profiles. Therefore, a direct point-to-point comparison of the sound-speed profiles obtained by different inversion techniques without taking into account differences in the sensitivity and averaging kernels, as presented for instance in \cite{Gizon2009}, is not justified.

\begin{figure}[h]
\begin{center}
\includegraphics[width=0.5\linewidth]{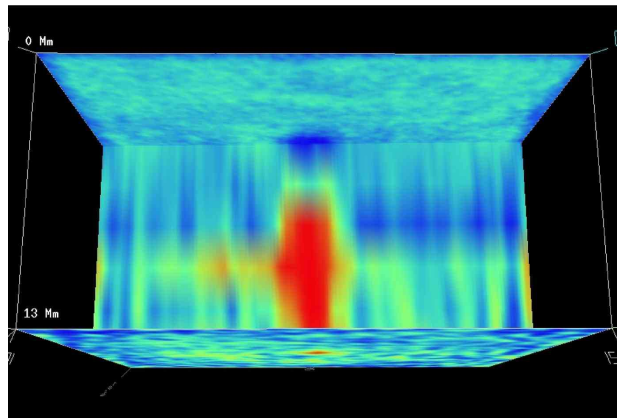}
\end{center}
\caption{\label{fig7}A vertical cut through the 3D wave-speed structure of AR 9787 obtained by the time-distance inversion technique. Red color shows positive variations; blue color shows negative variations.}
\end{figure}
\begin{figure}[h]
\begin{center}
\includegraphics[width=0.7\linewidth]{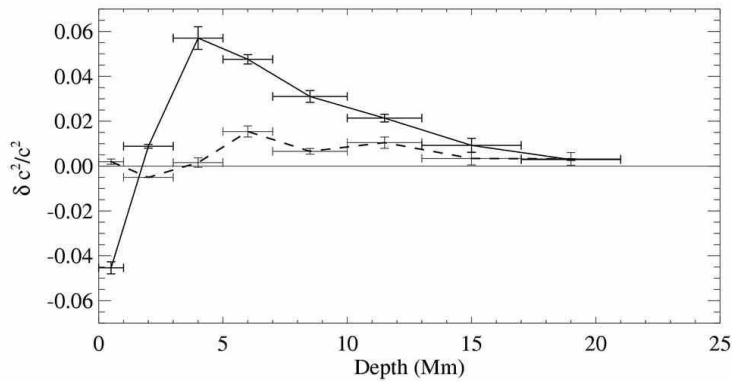}
\end{center}
\caption{\label{fig8}Relative variations of the seismic sound-speed structure with depth for AR 9787 obtained from the time-distance inversions: solid curve shows the results averaged over the sunspot; dotted curve shows the results of averaging for the 15-degree ring-diagram region, weighted with the local oscillation power. The horizontal bars show the width of the averaging layers in the time-distance inversion procedure \cite{Zhao2010a}. The error estimates are obtained from RMS variations calculated for three different 8-hour periods.}
\end{figure}

To demonstrate the importance of these differences we carried out analysis of AR 9787 by the time-distance helioseismology method, using the codes developed for the SDO/HMI data analysis pipeline \cite{Zhao2010a}. The time-distance inversion results are obtained in the ray-path approximation, and are shown in Fig.~\ref{fig7}. They reveal the typical two-layer seismic structure. In Figure~\ref{fig8}, we plot the depth dependence of the sound-speed variations averaged over the sunspot area (solid curve) and the variations averaged over the 15-degree ring-diagram analysis area (dashed curve). In the latter case, the sound-speed variations were averaged with weights proportional to the relative acoustic power distribution in this area. In this case, the sound-speed variation is much weaker, and also has a different depth dependence. In particular, the transition region from  the negative to positive variations and the maximum of the positive variation are shifted to deeper layers.

\begin{figure}[h]
\begin{center}
\includegraphics[width=0.7\linewidth]{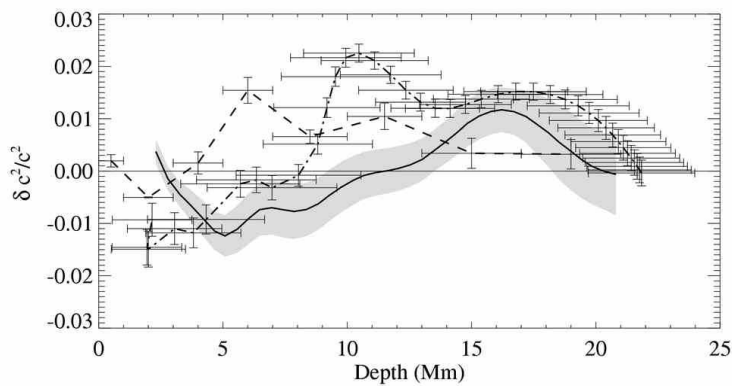}
\end{center}
\caption{\label{fig9}Comparison of the seismic sound-speed perturbations for AR 9787 obtained by the ring-diagram inversion techniques: SOLA (solid curve) and OLA (dash-dots) and by the power-weighted averaging of the time-distance inversion results (dashed curve).}
\end{figure}

In Figure~\ref{fig9}, we compare the ring-diagram inversion result obtained by the SOLA and OLA inversion methods and the power-weighted average of the time-distance inversion results. These results have similar amplitude variations and both show the two-layer structures. However, the transition between the negative and positive variations occurs at different depths: $\sim 4$~Mm for the time-distance result, and $\sim 5-8$~Mm for the ring-diagram inversions.
While formally this difference is within the resolution of the averaging kernels, the sound-speed structure obtained by the ring-diagram technique appears systematically more spread with depth than the structure obtained by the weighted averaging of the time-distance results.

It will be interesting to investigate this difference in more detail. We plan to continue this investigation for other active regions and also by using numerical wave simulations \cite{Hartlep2008,Parchevsky2009} for various models of the subsurface sound-speed and magnetic field structures.

\section{Conclusions}

Our investigation of the subsurface seismic structure of AR 9787 shows that the inversion results obtained by two different methods of local helioseismology, the ring-diagram analysis and time-distance helioseismology, are consistent with most of the previous results for other active regions, revealing the characteristic two-layer structure with a negative variation of the sound speed in a shallow subsurface layer and a positive variation in the deeper interior. However, there are significant quantitative differences between the inversion results obtained by the different techniques and different inversion methods. In particular, the seismic structure of the active region inferred by the ring-diagram method appears more spread with depth than the structure obtained from the time-distance technique.

In this paper, we point out that the quantitative comparison of the inversion results is not straightforward because of the substantially different spatial resolutions of the helioseismology methods. The quantitative comparison must take into account differences in the sensitivity and resolution. In particular, because of the acoustic power suppression the contribution of the sunspot seismic structure to the ring-diagram signal can be substantially reduced. We show that taking into account this effect  reduces the difference in the depth of the sound-speed transition region. In this analysis, we assumed that  the time-distance inversion results can be averaged over the ring-diagram analysis area, with weights proportional to the acoustic power distribution.  However, this assumption needs to be confirmed by numerical simulations. Further detailed analysis of the sensitivity, resolution and averaging properties is necessary for consolidation of the ring-diagram and time-distance inversion results.

 Our results obtained by the two local helioseismology methods indicate that the seismic structure of sunspots is probably rather deep, and extends to at least 20 Mm below the surface. If confirmed by further studies this conclusion has important implications for development of theoretical models of sunspots.

\paragraph{Acknowledgement.} We thank C.~Balder for providing the ring-diagram fitting data and inversion results obtained by the SOLA technique, and helpful discussions.

\providecommand{\newblock}{}

\end{document}